# Personalized Resource Allocation in Wireless Networks: An AI-Enabled and Big Data-Driven Multi-Objective Optimization

RAWAN ALKURD, (Member, IEEE), IBRAHIM Y. ABUALHAOL, (Senior Member, IEEE), AND HALIM YANIKOMEROGLU, (Fellow, IEEE)
Department of Systems and Computer Engineering, Carleton University, Ottawa, ON K1S 5B6, Canada

Corresponding author: Rawan Alkurd (rawan.alkurd@carleton.ca)

This work was supported by the Vanier Canada Graduate Scholarship.

**ABSTRACT** The design and optimization of wireless networks have mostly been based on strong mathematical and theoretical modeling. Nonetheless, as novel applications emerge in the era of 5G and beyond, unprecedented levels of complexity will be encountered in the design and optimization of the network. As a result, the use of Artificial Intelligence (AI) is envisioned for wireless network design and optimization due to the flexibility and adaptability it offers in solving extremely complex problems in real-time. One of the main future applications of AI is enabling user-level personalization for numerous use cases. AI will revolutionize the way we interact with computers in which computers will be able to sense commands and emotions from humans in a non-intrusive manner, making the entire process transparent to users. By leveraging this capability, and accelerated by the advances in computing technologies, wireless networks can be redesigned to enable the personalization of network services to the user level in real-time. While current wireless networks are being optimized to achieve a predefined set of quality requirements, the personalization technology advocated in this article is supported by an intelligent big data-driven layer designed to micro-manage the scarce network resources. This layer provides the intelligence required to decide the necessary service quality that achieves the target satisfaction level for each user. Due to its dynamic and flexible design, personalized networks are expected to achieve unprecedented improvements in optimizing two contradicting objectives in wireless networks: saving resources and improving user satisfaction levels. This article presents some foundational background on the proposed network personalization technology and its enablers. Then, an AI-enabled big data-driven surrogate-assisted multi-objective optimization formulation is proposed and tested to illustrate the feasibility and prominence of this technology.

## I. INTRODUCTION

Over the past decade, the convergence of Internet of Things (IoT) and Ambient Intelligence (AmI) technologies have paved the way for more connected, adaptive, proactive, and smart environments. Nonetheless, although adaptive environments are designed to support people in their daily life, human emotions and preferences are rarely taken into considerations. Nowadays, users interact with technology through two senses, sight and sound. However, by 2025, it is envisioned that advanced technology will enable a full internet of senses, including touch, taste, smell, and mind. Also, by 2030, communicating thoughts digitally could be possible, which in effect, will replace the current user interfaces, such as mouses and keyboards, by our brains [1]. The emerging internet of senses technology will enable the transparent (i.e., without direct user feedback) integration of human emotions and preferences, which will improve the personalized user experience for various services and products.

In wireless networks, emotion-aware applications have proved to offer better user experience and improved systems efficiency. Examples of such applications include cognitive

  

radio ad-hoc networks [2] and mobile cloud computing [3]. One of the main emerging emotion-aware applications in wireless networks is network personalization, in which user experience is greatly enhanced by providing services personalized to users' individual needs and expectations in continuously varying contexts.

Unlike personalized networks, current wireless networks are over-provisioned to unnecessarily provide high Quality-of-Service (QoS) levels in order to achieve high satisfaction levels for all users. In certain contexts, some users may have lower QoS requirements, yet the network will always attempt to provide higher QoS levels, and consequently charges users more for the unnecessary high-quality services. Arguably, although this non-granular average-based single-objective approach is currently adopted by all operators, it is far from optimum and it is costing the majority of users more money for the provided extra bandwidth they do not need or use. Besides, this over-provisioned design will not be able to cope with the emerging network requirements as future wireless networks are designed to support the emerging bandwidth-hungry applications, such as Virtual Reality (VR), Augmented Reality (AR), and self-driving cars. Tackling the exploding rate demand issue by continuously investing in new infrastructure will eventually make wireless networks unprofitable or make network services very expensive. Therefore, there is a tremendous need to efficiently utilize scarce resources already available in wireless networks. Furthermore, the emerging wireless network applications require network services to be delivered with a variety of network performance characteristics (e.g., rate, latency, security, and Quality-of-Experience (QoE)), which poses fundamental technical challenges for the management of user experience. By contrast, personalized wireless networks are envisioned to micro-manage resources in a way that meets the expectations of each user in the network while using a minimum amount of resources. This will provide operators with improved flexibility of operation in terms of the amounts of consumed resources and personalized user satisfaction (rather than one averaged satisfaction value assumed to be good for most users). Also, enabling networks to make more personalized decisions (e.g., configurations) and optimized actions (e.g., resource allocation) is crucial and indispensable to achieving the ultimate balance between network resources and user satisfaction. Finally, wireless networks are considered the most inefficient systems in terms of energy consumption. Personalizing service quality instead of overflowing the network with unneeded high amounts of network resources could contribute to the reduction of energy consumption; hence decrease toxic emissions.

The main contribution of this article revolves around addressing the personalized decision-making process that is responsible for making optimized, fine-grained, and personalized actions in wireless networks. The decision process in personalized wireless networks is based on the intelligence created by Machine Learning (ML) engines. The primary use of ML in personalized networks is to build surrogate models for user satisfaction behavior, which is highly dynamic and continuously evolving. This article begins by presenting some fundamental concepts associated with personalized wireless networks. Then, of the various decision-making processes in wireless networks, we shed light on the personalization of the resource allocation process. The main premise of personalized resource allocation is to achieve optimum allocation such that maximum user satisfaction levels are achieved using the minimum amount of resources. To this end, personalized wireless networks should be designed to optimize two correlated and contradicting objectives in real-time: user satisfaction and resource utilization. We term the described optimization problem by Optimum Personalized Resource Allocation (OPA) problem.

In this article, we model OPA as a Multi-Objective Optimization (MOO) problem. In wireless networks, decisions are made in real-time; hence, to maintain the proactivity of the network, the optimization process and decisions are also required to be in real-time. As a result, although exact optimization algorithms that are based on mathematical programming produce the best possible solutions, they are slow and computationally expensive; hence not feasible due to the complexity of OPA. Instead, to speed up the optimization process, we resort to data-driven evolutionary optimization to approximate the Pareto front solutions. In view of this, we present a review and categorization of data-driven Evolutionary MOO (EMOO) followed by a discussion on the benefits and challenges of employing EMOO in personalized wireless networks. Another important aspect of the problem is integrating user satisfaction behavior into the optimization process. To actualize this in real-time, we utilize a surrogate model to approximate the personalized user satisfaction behavior of network users. The proposed surrogate model is ML-based and built using Deep Neural Networks (DNNs). Then, in order to maintain and manage the surrogate models, we propose a surrogate-management framework that employs the collection of select user satisfaction feedback measurements in real-time. The proposed framework reduces the risk of solutions divergence and the effect of uncertainty introduced by surrogate models. Besides, it is designed to continuously enhance the performance of the surrogate models as more data arrive in the network. Afterward, we formulate and solve the OPA problem using EMOO. Through several experiments, we analyze the optimum Pareto front solutions for various EMOO algorithms. Then, using the best algorithm, we compare the personalized and non-personalized networks in terms of the amount of saved resources and user satisfaction levels in the network. Moreover, we study the effect of uncertainty introduced by the surrogate models on the quality of the produced Pareto front solutions. Finally, we conduct a scalability analysis to explore the effect of higher numbers of users and the effect of varying the Number of Function Evaluations (NFEs) on the performance of the simulated algorithms and the quality of solutions.

## II. WIRELESS NETWORK PERSONALIZATION: CHALLENGES AND SOLUTIONS

To realize wireless network personalization, certain design and implementation-related challenges need to be overcome. This section summaries these challenges and discusses the proposed solutions.

### A. INTEGRATION INTO WIRELESS NETWORKS

The first design challenge of personalized networks is *the integration of network personalization into current wireless networks*. Since wireless networks are already complex and highly structured systems, as shown in Fig. 1, our vision is to consolidate an intelligent layer into wireless network layers dedicated to personalizing network decisions. The personalization layer is responsible for digesting and analyzing data, modeling complex and dynamic user behavior using ML, and utilizing the created intelligence in making optimized and personalized network decisions. This design enables wireless network personalization to act as an orthogonal system that can be supported in any wireless networks, and hence reduce complexity. Also, this type of modularity will enable the personalization of diverse sets of applications and problems in networks. To address the integrability issue, we propose in [4] a framework in which we illustrate how personalization can be integrated into current wireless networks in such a way that enables the coherent operation of both systems with reduced complexity. The framework describes the process of data collection, processing, and the process of utilizing user satisfaction behavior information to learn, predict, and optimize based on user needs and expectations in a certain context.

### B. MEASURING USER SATISFACTION

The second design challenge of wireless network personalization is to find a way *to quantify and measure user satisfaction in wireless networks*. The importance of this challenge stems from the fact that personalized networks require the continuous measurement and tracking of user satisfaction. In wireless networks, user satisfaction is highly subjective, complicated, and changes dynamically depending on various factors. For this reason, mathematical expressions that attempt to model the relationship between user satisfaction and other factors do not yield accurate results. Therefore, adopting data-driven approaches that are backed by ML and AI techniques are the best strategy to model and predict user satisfaction in wireless networks. Nonetheless, due to the lack of a dynamic user satisfaction model, researchers and service providers were not able to dynamically quantify and predict the real-time personalized satisfaction behavior of users in wireless networks. To this end, we proposed in [4] a dynamic user satisfaction model that is based on the notion of Zone of Tolerance (ZoT). As shown in Fig. 2, we propose dividing user satisfaction into levels, where each level is associated with a certain range of QoS. The division and number of satisfaction levels could vary depending on service providers' preferences. In order to achieve a satisfaction level $i$, the user should receive a QoS within $ZoT_i$. Although QoS seems the main factor influencing user satisfaction in wireless networks, the gap between demanded QoS ($QoS_d$) and provided QoS ($QoS_p$) is what actually does. To incorporate this gap into our satisfaction model, as shown in Fig. 2, we define the variable $\Delta$, which refers to the difference between the demanded and provided QoS ($QoS_d$-$QoS_p$). Also, the minimum (adequate) QoS required to achieve a satisfaction level $i$ is referred to as $QoS_{a_i}$. In our proposed ZoT model, each user satisfaction behavior (i.e., the relation between $\Delta$ and satisfaction) is associated with a certain context. Context is a multi-variate variable composed of various context variables, such as time, location, and speed. It is worth mentioning that QoS can be a vector with several elements, such as rate, reliability, latency, and jitter. Nonetheless, for simplicity, we assume that QoS is solely defined by rate.

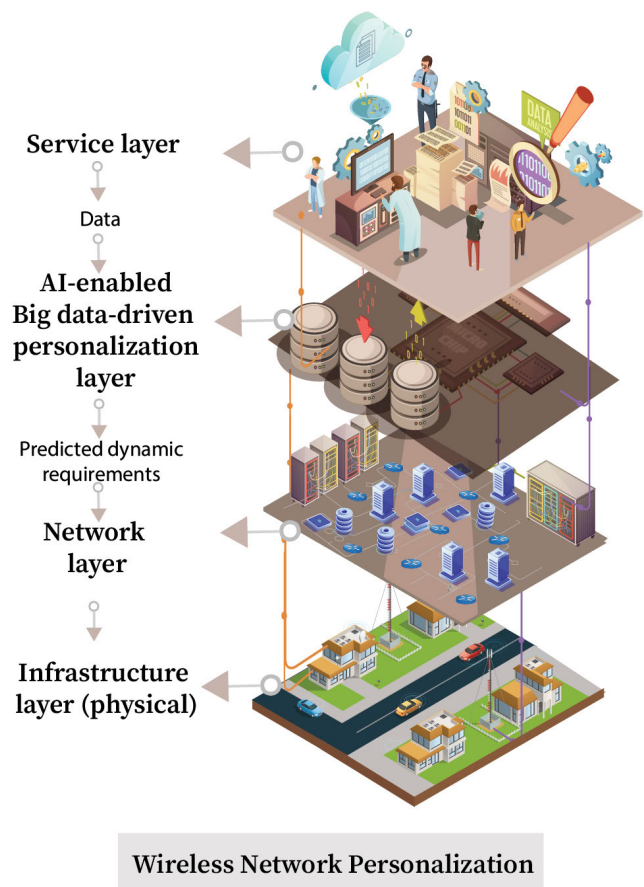

**FIGURE 1.** Wireless network personalization: a big data-driven AI-based layer.

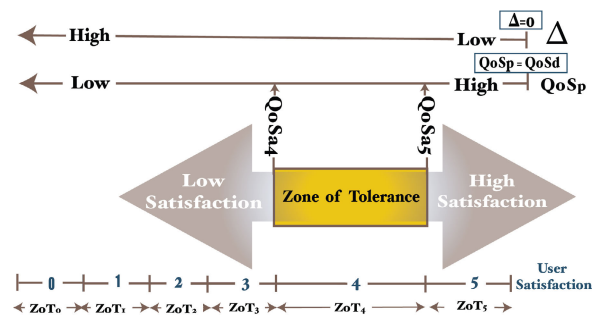

**FIGURE 2.** The proposed satisfaction Zone of Tolerance model [4].

### C. DATA COLLECTION

In addition to the aforementioned challenges, given that AI is a key component in wireless network personalization, data is an essential requirement. Therefore, understanding *the types of data that can be mined to personalized wireless networks* is cardinal for the successful integration of this technology. However, the lack of publicly available user data due to confidentiality and privacy concerns is slowing down the research and development in wireless network personalization and holding back innovation into new approaches for monitoring and predicting user satisfaction. Therefore, we proposed in [5], [6] a synthetic dataset structure designed based on Bayesian networks as well as Markov chain models. The proposed synthetic user behavior-satisfaction datasets can be utilized for data-driven user satisfaction prediction and optimization from context information. The synthesized datasets are meticulously designed to have realistic characteristics and personas, and therefore behave in the same manner as a real user behavior dataset [7]. In this article, we utilize the datasets proposed in [6] and [5] to solve the OPA problem, which will be discussed later in the paper. The datasets are publicly available in a GitHub repository [8]. Datasets in [8] are designed for four distinct personas. For the purpose of this article, we will work with the Working Professional Persona (WPP) dataset. Table 1 shows the features of the WPP dataset and an example of their values.[1]

### D. OPTIMUM DECISION MAKING

Finally, one of the most fundamental design challenges that need to be addressed is *the process of integrating the optimized decisions made by the data-driven intelligent personalization layer with network decisions in order to make relevant personalized actions*. The rest of this article focuses on addressing this challenging aspect of personalized wireless networks. Undoubtedly, there are several other design issues as well as implementation-related challenges that need to be addressed in order to realize wireless network personalization. Nonetheless, for the purpose of this article, we limit our discussion to the aforementioned points.

## III. DATA-DRIVEN SURROGATE-ASSISTED EVOLUTIONARY MULTI-OBJECTIVE OPTIMIZATION

Many problems in all sorts of research fields are formulated as optimization problems. While optimization problems were traditionally approached using mathematical programming, the complexity level of current problems have led researchers in academia and industry to move towards more heuristics/metaheuristics approaches. In contrast to mathematical programming, heuristics/metaheuristics optimization algorithms are less sensitive to the formulation of the optimization problem. This is considerably important for wireless network optimization problems due to their scale and complexity level. Generally, heuristics/metaheuristics optimization algorithms are of two main classes, Evolutionary Algorithms (EAs) [9] and Swarm Intelligence-based Optimization Algorithms (SIOAs) [10]. In this article, we employ EAs to solve the proposed and formulated optimization problem.

TABLE 1. Features of the WPP dataset.

| No. | Feature name | Units | Example |
|---|---|---|---|
| 0 | Date | - | 2018-01-10 |
| 1 | Time | - | 14:55:02 |
| 2 | Day | - | [HTML]000000 Wednesday |
| 3 | Classified days | - | Weekday |
| 4 | Time period | - | Afternoon |
| 5 | Location | - | [22, 73] |
| 6 | Location name | - | work |
| 7 | Speed | km/hr | 5.2 |
| 8 | Speed range | - | low |
| 9 | Activity | - | Walking |
| 10 | Request arrived | - | 1 |
| 11 | Application | - | WhatsApp |
| 12 | Service | - | Picture |
| 13 | Demand rate | kbps | 867 |
| 14 | Min rate | kbps | 600 |
| 15 | Given rate | kbps | 802 |
| 16 | $\Delta$ | - | 65 |
| 17 | Max $\Delta$ | - | 267 |
| 18 | Satisfaction | - | 4.0 |

### A. EVOLUTIONARY OPTIMIZATION ALGORITHMS

EAs are a class of metaheuristics population-based optimization algorithms, where multiple candidate solutions are maintained in parallel. EAs are designed based on the idea of the survival of the "fittest" solution in order to evolve a population that is a good approximation of the global optimum that we wish to find [11]. The fitness of an evolved solution is a measure of its quality at solving the problem. Block 1 in Fig. 3 illustrates the process cycle of evolutionary computation. At each cycle, EAs begin with generating parents (populations of candidate solutions). Then, offspring solutions are generated using various variation operations, such as crossover and mutation. Lastly, in order to select the parent solution for the next cycle, the quality (or fitness) of the generated offspring solutions are evaluated using the objectives and constraints.

There are several advantages of EAs that drive researchers to utilize them for solving various optimization problems,

---
[1] The dataset in [8] has other features, such as real sensor measurements. However, for the purpose of this article, we consider only the features listed in Table 1.

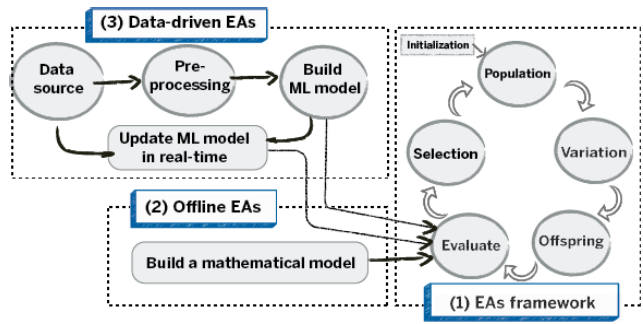

**FIGURE 3.** A comparison between data-driven (online) and offline EAs frameworks.

of which the most important is that they do not necessitate analytical modeling and formulation of the objectives and constraints functions associated with the optimization problem. This is considered beneficial since many emerging problems, including the proposed OPA problem, cannot be expressed using mathematical modeling. The mathematical formulation could require the integration of many relaxations into the problem, which, in effect, will limit the practicality and accuracy of the produced and solved problems. Another important advantage of EAs is that they can operate with little domain knowledge, which creates more robust models that are less susceptible to the various mathematical features of problems, such as convexities and discontinuities. Another advantage that adds to the robustness of EAs is that they are based on stochastic search, which samples the entire population making them less vulnerable to local optimums.

The advantages of heuristics/metaheuristics optimization come at a price. Since EAs rely on iterating the evaluation process of the objective and constraints associated with the solutions population, the higher the degree of accuracy required, the higher the computational power required. However, supercomputers and distributed computing technologies are improving rapidly to the point that the use of big data analytics and EAs for practical near-real-time applications are currently possible. As a result, EAs have been utilized in many fields and applications, such as automated data mining to extract meaningful information [12], finance and economics [13], and wireless sensor networks [14].

### B. ONLINE vs. OFFLINE OPTIMIZATION

Although EAs do not require the analytical and mathematical formulation of objectives and constraints, most EAs in the literature assume the availability of such models. Optimization algorithms that assume the availability of mathematical models to assess objectives and constraints are referred to as offline optimization algorithms. Unfortunately, offline optimization is not a valid approach for many dynamic problems that involve rapidly changing features, requirements, and behaviors. In comparison, online optimization continuously digests data from the problem's environment to make decisions based on updated information flowing to the algorithm, making the optimization process more dynamic and adaptive.

EAs that are based on data are called data-driven evolutionary optimization algorithms. As shown in Fig. 3, the main difference between offline EAs (Block 2, Fig 3) and data-driven EAs (Block 3, Fig 3) is the method used in the evaluation process. Offline EAs utilize analytical objectives and constraints to evaluate the fitness of solutions, whereas data-driven EAs utilize data-driven models. The proposed OPA problem in this article utilizes a data-driven model to estimate real-time user satisfaction in networks; hence, data-driven (online) EAs are utilized to find the optimum decisions.

### C. INTERACTIVE EVOLUTIONARY COMPUTATION

Interactive Evolutionary Computation (IEC) is evolutionary computation applied to optimize systems based on humans' subjective opinions and expectations [15]. The IEC technology embeds a user in the optimization system in which the user is considered to be a black box. There are several reasons for integrating user input with particularly EAs as opposed to other optimization methods. One of the most prominent reasons is that EAs do not require the use of gradient information to search the space, which in most scenarios cannot be computed for such problems. Examples of problems that capitalize on IEC algorithms include mental health measurement [16] and emotional music generation [17].

Although the decisions in personalized wireless networks are optimized based on users' subjective opinions and expectations, users are not actively logging their satisfaction levels. Instead, as shown in Fig. 4, user satisfaction is captured in a non-intrusive manner from sensors data using AI. Also, the trained user satisfaction ML engine models user behavior and expectations, which enables the network to repeatedly evaluate user satisfaction in the optimization process in a non-intrusive manner and in a relatively short time. Considering the fact users are involved in the personalized optimization process, the proposed OPA problem in this article can be considered as an IEC problem. In the literature, many researchers proposed solutions to improve the performance and efficiency of IEC problems. For example, preference-based (or progressive) interactive evolutionary optimization reduces the required numbers of function evaluations by involving the decision-maker in the intermediate generations of the algorithm; hence focus computations on the targeted Pareto front solutions [18], [19]. Even though IEC-based algorithms have several benefits and can greatly enhance the performance of optimization algorithms, for the purpose of this article, we employ well-known Multi-Objective Evolutionary Algorithms (MOEAs) that are not necessarily optimized for IEC problems.

## IV. SURROGATES IN PERSONALIZED WIRLESS NETWORKS

The micromanagement and personalization of wireless networks require the continuous tracking and measurement of personalized user satisfaction behavior for all users. Such a level of granularity and dynamic behavior modeling cannot be achieved using the traditional mathematical models or the average-based ML models proposed in the literature [20]–[23]. Therefore, in [4], we proposed the utilization of personalized ML models designed based on the proposed user ZoT model. Besides, the ML models are optimized based on context data collected by the network. These personalized satisfaction behavior models are employed by personalized networks to personalize the allocation of resources based on the satisfaction behavior of each user.

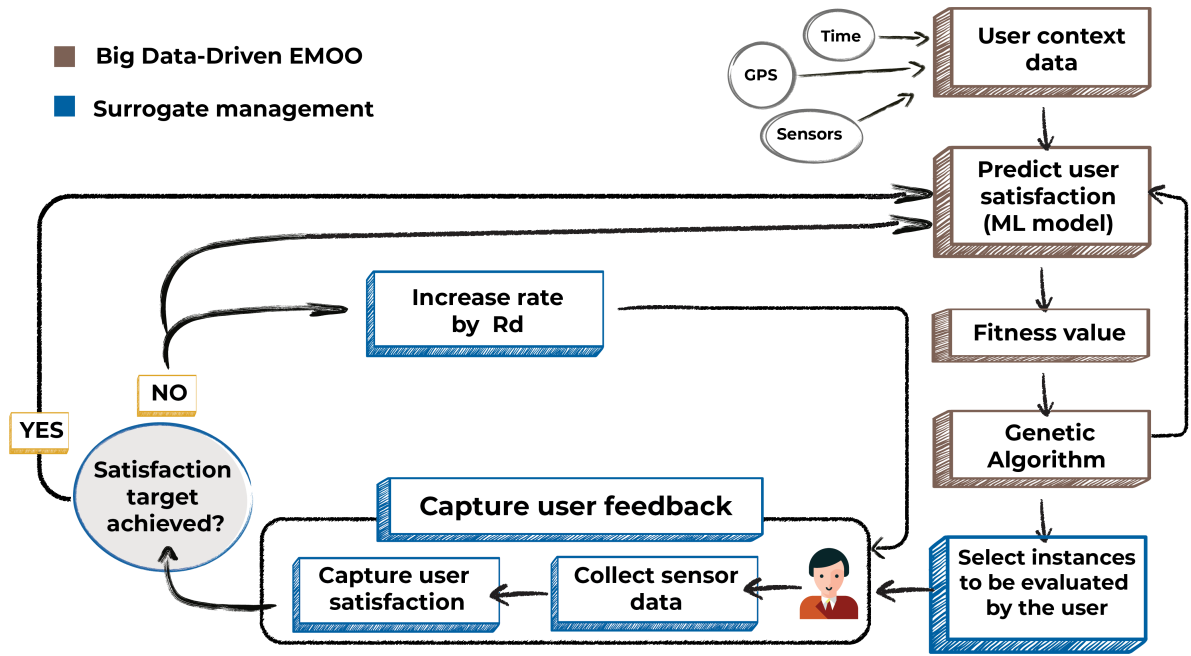

**FIGURE 4.** EMOO management framework to optimize personalized wireless networks.

### A. WHY SURROGATES ARE ADVANTAGEOUS?

In the optimization literature, ML models that are used to estimate the relationship between the inputs and outputs of systems are known as ML-based surrogate models (or meta-models). In this section, we discuss the main drivers for adopting surrogate models in personalized wireless networks.

#### 1) REAL-TIME USER SATISFACTION EVALUATION

Although user satisfaction behavior in wireless networks cannot be modeled analytically, the network can still capture the satisfaction measurements from users. Although these measurements may not be done in real-time, they can act as the exact function evaluations (or reference evaluations) for the evolutionary optimization algorithm. The problem with measuring user satisfaction is the associated cost. Collecting data and extracting user satisfaction information is expensive and time-consuming. Due to the nature of wireless network applications, wireless networks are designed to take decisions in real-time. Therefore, for practical reasons, real-time network decisions should not be designed to be dependent on the captured user satisfaction. Evolutionary optimization of wireless network problems will not be able to afford thousands of real-time evaluations required by EAs. Instead, surrogate models are comparatively faster and rely on historical data and user patterns to predict user satisfaction values in real-time. Accordingly, surrogate models are considered an essential enabler for personalized wireless networks.

#### 2) DYNAMICALLY EVOLVING USER SATISFACTION MODELS

Our expectations of wireless networks dynamically change with several factors that constitute multi-variate contexts. Also, the speed of behavioral changes can be in the order of seconds, which imposes another limitation on the modeling process of user behavior in wireless networks. Although users could have repeating patterns in their short-term behavior and expectations, their long-term behavior and satisfaction patterns can change. This can be explained by different factors, such as age, professional development, career type, etc., which are usually accompanied by interest changes. Since the surrogate models are built using ML models, and due to the fact that such surrogates are continuously trained and validated through user feedback data, they are able to capture the short-term changes in user behavior patterns as well the behavioral changes over the long run. This ensures the dynamic design and adaptivity of the network.

### B. MANAGMENT OF SURROGATES IN PERSONALIZED NETWORKS

Management of surrogates, which involves the process of using and updating the models, plays an essential role in maintaining acceptable performance of surrogate-assisted optimization [24]. Generally, surrogate models are assumed to be of high fidelity; hence, the exact fitness functions are not utilized in the evolutionary optimization computation. Nevertheless, this approach can run the risk of convergence to incorrect solutions [25]. To address this issue, many strategies for managing surrogate models using exact fitness functions are proposed in the literature [24]. Nonetheless, since the OPA problem proposed in this article does not have an exact fitness function, the implemented MOEAs are solely dependent on the approximated ML-based surrogates. Hence, to manage user satisfaction surrogates in personalized wireless networks, we adopt a different approach. Fig. 4 illustrates the proposed framework for solving the OPA problem using evolutionary computation. As shown in Fig. 4, the proposed framework essentially relies on the trained ML-based user satisfaction surrogates to compute the satisfaction fitness values during the optimization process. However, even though the exact fitness function for user satisfaction behavior in personalized wireless networks does not exist, user satisfaction feedback can be measured and utilized to enhance the accuracy of the implemented surrogates and prevent the network from converging to inaccurate solutions. As shown in Fig. 4, user satisfaction feedback is measured and fed back to the surrogate model in order to actively validate and correct inaccurate solutions produced by the optimizer. Also, the surrogate model uses the continuously arriving data samples to learn and enhance its performance. Moreover, when the measured user satisfaction levels do not match the optimized targeted levels by the optimizer, as illustrated in Fig. 4, the proposed framework suggests to gradually increase/decrease the provided resources to the user by $R_d$ while continuously measuring the actual user satisfaction levels. Then, the collected user satisfaction behavior data at that particular context is used to enhance the performance of the approximated ML-based satisfaction surrogate. This process is critical as it prevents wireless networks from continuously providing services to users with satisfaction levels that do not meet the service providers' standards and requirements.

Notably, only a small number of instances are re-evaluated and validated by the real user satisfaction behavior in the network. There are several algorithms and solutions proposed in the literature for choosing such individuals or instances, including selecting solutions with high uncertainty levels [26], [27] and choosing representative and good solutions [28]–[30]. The implementation and integration of these strategies into our proposed framework in Fig. 4 are out of the scope of this article.

### C. SURROGATE MODEL DESIGN AND PERFORMANCE

Various ML models can be utilized to build surrogates, including linear models, support vector machines [31], and Gaussian processes [32]. In this article, we adopt DNNs to capture the complicated patterns that exist within the collected user data. There are several advantages of employing DNNs to model and predict user satisfaction levels in wireless networks of which the most important is their scalability and ability to automate feature extraction from data that has complex structures and correlations. Since building accurate and complex ML engines to predict user satisfaction in wireless networks is not the main focus of this article, we adopt a simple DNN model with four layers as follows:
- First hidden layer (Layer 2): 128 neurons.
- Second hidden layer (Layer 3): 32 neurons.
- Third hidden layer (Layer 4): 16 neurons.
- Fourth hidden layer (Layer 5): 8 neurons.

The data fed into the model is pre-processed using several steps including scaling, encoding, and balancing. In [7], we discuss the details of the implemented preprocessing steps for the personalized networks dataset utilized in this article. To examine the performance of the utilized DNN model, we implement a 10-folds cross-validation test. Tabel 2 summarizes the performance of the implemented ML model.

**TABLE 2.** Performance of the adopted DNN design in terms of accuracy.

| Accuracy of individual folds in % | | | | | | | | | |
|---|---|---|---|---|---|---|---|---|---|
| fold-1 | fold-2 | fold-3 | fold-4 | fold-5 | fold-6 | fold-7 | fold-8 | fold-9 | fold-10 |
| 95.45 | 95.76 | 94.61 | 95.66 | 95.64 | 94.93 | 95.46 | 95.54 | 95.30 | 95.41 |
| Average Accuracy | | | | | Std. of Accuracy | | | | |
| 95.38 % | | | | | 0.34 % | | | | |

## V. DATA-DRIVEN MULTI-OBJECTIVE OPTIMIZATION OF RESOURCES IN PERSONALIZED WIRELESS NETWORKS

In order to study the benefits of integrating personalization into wireless networks, we model, formulate, and solve the resource allocation problem for personalized wireless cellular networks (i.e., OPA). As shown in Fig. 5, the resource allocation algorithm for personalized wireless networks accepts two groups of inputs, user context values, and network/system context values. User context is a set of variables that affect user satisfaction behavior in the network. As shown in Fig. 5, examples of user context variables include user ID, time, location, speed, application, and $QoS_d$. On the other hand, network context is the set of network variables that affect network conditions, such as noise power, channel gain, Signal to Noise Ratio (SNR), packet rate, and throughput. Taking into consideration user and network context, each user is assigned a set of Resource Blocks (RBs) determined based on the predicted user satisfaction behavior at each particular instant. The resource allocation algorithm will output the optimum set of RBs for each user such that the required satisfaction level is achieved using the minimum amount of resources.

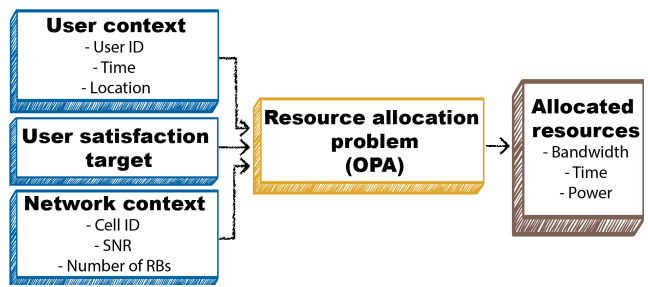

**FIGURE 5.** Inputs and outputs of the Optimum Personalized Resource Allocation (OPA) problem for wireless networks.

### A. PROBLEM DESCRIPTION
Usually, resource scheduling problems for networks are modeled as an optimization problem with a single objective that maximizes throughput, spectral efficiency, or fairness under certain constraints. Besides, the objective could be a trade-off between throughput/spectral efficiency and fairness [33]. In this article, OPA is modeled as a MOO problem. MOO problems are used to model optimization problems with more than one conflicting objective. OPA is defined as a bi-objective optimization problem that maximizes two conflicting objective functions: the total $\Delta$ (QoS$_d$-QoS$_p$) for all users, which represents the amount of saving in network resources, and the average satisfaction for all users. Typically, MOO problems are solved by finding the set of mutually nondominant solutions called the Pareto front. In a Pareto front solution set, there is no solution better than the other as all solutions trade off the conflicting objective functions [34]. Using MOO, along with the available satisfaction information in personalized networks, the network can make granular personalized resource allocation decisions for each user to ensure that the required satisfaction level is achieved with the minimum cost (i.e., resources).

### B. PROBLEM FORMULATION
#### 1) SYSTEM MODEL
let $\mathcal{B} = \{1, 2, \ldots, B\}$ eNBs available in the network. The number of user equipment connected to eNB $b$ is denoted by $\mathcal{U}_b = \{u_{(1,b)}, u_{(2,b)}, \ldots, u_{(I_b,b)}\}$, where $I_b$ is the number of users connected to eNB $b$. Without loss of generality, we assume that $I_b$ is constant for all $b \in \mathcal{B}$. For each eNB $b \in \mathcal{B}$, the number of available RBs is denoted by $\mathcal{N} = \{n_1, n_2, \ldots, N\}$. We define $h_{b,u_b}^{(n)}$, where $u_b \in \mathcal{U}_b$, as the link gain between eNB $b$ and $u_b$ over RB $n$. The power Signal to Interference Noise Ratio (SINR) between eNB $b$ and $u_b$ over RB $n$ is as follows:

$$\gamma_{b,u_b}^{(n)} = \frac{P_{b,u_b}^{(n)} h_{b,u_b}^{(n)}}{\sum\limits_{\forall u_b \in \mathcal{U}_b, j \neq b, j \in \mathcal{B}}^{n} P_{b,u_b}^{(n)} h_{b,u_j}^{(n)} + N_0 B_{RB}}, \quad (1)$$

where $P_{b,u_b}^{(n)}$ is the power assigned for the link between $b$ and $u_b$ over RB $n$ for all $b \in \mathcal{B}$, $N_0$ denotes the thermal noise, and $B_{RB}$ is the resource block bandwidth. The power allocation vector for user $i$ connected to eNB $b$ is $\mathcal{P}_{b,u_b} = [P_{b,u_b}^{(1)}, P_{b,u_b}^{(2)}, \ldots, P_{b,u_b}^{(N)}]$. For simplicity, we assume that interference between cells is negligible and SINR for all $b \in \mathcal{B}$ is reduced to the following SNR:

$$\gamma_{b,u_b}^{(n)} = \frac{P_{b,u_b}^{(n)} h_{b,u_b}^{(n)}}{N_0 B_{RB}}. \quad (2)$$

The achievable downlink data rate for all $u_b \in \mathcal{U}_b$ over RB $n$ is given by

$$r_{b,u_b}^{(n)} = B_{RB} \log_2(1 + \gamma_{b,u_b}^{(n)}). \quad (3)$$

#### 2) Solution ENCODING: RESOURCE BLOCK AND POWER Allocation
The RB allocation indicator is denoted by binary decision variable $x_{u_b}^{(n)} \in \{0, 1\}$, where

$$x_{u_b}^{(n)} = \begin{cases} 1, & \text{if RB } n \text{ is assigned to } u_b, \\ 0, & \text{otherwise.} \end{cases} \quad (4)$$

The RB allocation vector for user $u_b$ connected to eNB $b$ is $\mathcal{X}_{u_b} = [x_{u_b}^{(1)}, x_{u_b}^{(2)}, \ldots, x_{u_b}^{(N)}]$. In each frame, the total rate achieved by $u_b$ over the allocated RBs is given by

$$\mathcal{R}_{u_b} = \sum_{n=1}^{N} x_{u_b}^{(n)} r_{b,u_b}^{(n)}, \quad (5)$$

for all $u_b \in \mathcal{U}_b$. Consequently, $\Delta_{u_b}$, which refers to the amount of saved resources by user $u_b$, is given by

$$\Delta_{u_b} = \mathcal{D}_{u_b} - \mathcal{R}_{u_b}, \quad (6)$$

where $\mathcal{D}_{u_b}$ is the rate demanded by user $u_b$ connected to eNB $b$. The sum of $\Delta_{u_b}$ for all $u_b \in \mathcal{U}_b$ is given by

$$\Delta_{\mathcal{U}_b} = \sum_{u_b \in \mathcal{U}_b} \mathcal{D}_{u_b} - \mathcal{R}_{u_b}. \quad (7)$$

The total rate achieved by all users connected to eNB $b$ is given by

$$\mathcal{R}_b = \sum_{u_b \in \mathcal{U}_b} \mathcal{R}_{u_b}. \quad (8)$$

#### 3) DATA DRIVEN OPTIMIZATION
Let $A_{u_b}$ be a $J$-dimensional feature space for $u_b$, where $A_{u_b} = [a_{(1,u_b)}, a_{(2,u_b)}, \ldots, a_{(J,u_b)}]$ and $a_{(j,u_b)}$ is the value of feature $j$ for user $u_b$. As mentioned earlier, since personalized satisfaction is predicted using the data collected from users, the output of a satisfaction level $\mathcal{S}_{u_b}$ is predicted using the deployed and tuned ML-based surrogate model. The inputs for the surrogate model are user ID, context features $A_{u_b}$, and user demand $\mathcal{D}_{u_b}$.

#### 4) OPTIMIZATION PROBLEM FORMULATION
In order to solve OPA, we formulate a MOO problem with two objectives: 1) maximize the average $\Delta_{\mathcal{U}_b}$ (i.e., amount of saved resources) for all users and 2) maximize the average satisfaction for all users. The formulated MOO has two decision variables: $\mathcal{X}_{u_b}$ and $\mathcal{P}_{u_b}$. Each objective function is associated with a set of constrained and formulated as follows:

  - **Maximize the average $\Delta_{\mathcal{U}_b}$ for all users:**

$$\max_{\mathcal{X}_{u_b}, \mathcal{P}_{u_b}} \frac{1}{I_b} \sum_{u_b \in \mathcal{U}_b} (\mathcal{D}_{u_b} - \sum_{n=1}^{N} x_{u_b}^{(n)} r_{b,u_b}^{(n)}), \quad (9a)$$

$$\text{s.t.} \sum_{u_b \in \mathcal{U}_b} x_{u_b}^{(n)} \leqslant 1, \quad \forall n \in N \tag{9b}$$

$$\sum_{u_b \in \mathcal{U}_b} \sum_{n=1}^{N} x_{u_b}^n P_{b,u_b}^n \leqslant \mathcal{P}_b^{\max} \tag{9c}$$

$$\mathcal{R}_{u_b,b} \leqslant \mathcal{D}_{u_b}, \quad \forall u_b \in \mathcal{U}_b \tag{9d}$$

$$P_{b,u_b}^{(n)} \geqslant 0, \quad \forall n \in N, u_b \in \mathcal{U}_b. \tag{9e}$$

- **Maximize the average satisfaction for all users:**

$$\max_{\mathcal{X}_{u_b}, \mathcal{P}_{u_b}} \frac{1}{I_b} \sum_{u_b \in \mathcal{U}_b} \mathcal{S}_{u_b}(A_{u_b}, \tag{10a}$$

$$(\mathcal{D}_{u_b} - \sum_{n=1}^{\mathcal{N}} x_{u_b}^{(n)} r_{b,u_b}^{(n)}(P_{b,u_b}^{(n)}))), \tag{10b}$$

$$\text{s.t.} \sum_{u_b \in \mathcal{U}_b} x_{u_b}^{(n)} \leqslant 1, \quad \forall n \in N \tag{10c}$$

$$\sum_{u_b \in \mathcal{U}_b} \sum_{n=1}^{N} x_{u_b}^n P_{b,u_b}^n \leqslant \mathcal{P}_b^{\max} \tag{10d}$$

$$\mathcal{R}_{u_b,b} \leqslant \mathcal{D}_{u_b}, \quad \forall u_b \in \mathcal{U}_b \tag{10e}$$

$$P_{b,u_b}^n \geqslant 0, \quad \forall n \in N, u_b \in \mathcal{U}_b \tag{10f}$$

$$\mathcal{S}_{u_b} \geqslant S_{\min,u_b}, \quad \forall u_b \in \mathcal{U}_b. \tag{10g}$$

The first objective function in (9a) maximizes the average $\Delta_{\mathcal{U}_b}$ ($\bar{\Delta}_{\mathcal{U}_b}$) in order to maximize resource-saving in the network. On the other hand, the second objective function in (10a) maximizes the average satisfaction for all users. Both objective functions contradict each other; hence, the solution set is expected to be a Pareto front, where the optimum points trade-off both objectives.

As for constraints, the first objective function (9a) has four constraints, of which constraint (9b) ensures that each RB is being used by no more than one user during a single instance. Also, the second constraint (9c) prevents each eNB from allocating a total power more than the budget power $\mathcal{P}_b^{\max}$. Besides, the third constraint (9d) limits the rate provided to each user to values less than the demanded rate $\mathcal{R}_{u_b,b}$. The fourth constraint (9e) ensures that the allocated power for each user $P_{b,u_b}^n$ is a positive value. On the other hand, the second objective function (10a) has five constraints. Constraints (10c), (10d), (10e), and (10f) are similar to the constraints associated with the objective function in (9a). The last constraint (10g) maintains a minimum satisfaction specified for each user. This constraint is added to differentiate among the targeted satisfaction levels for different users; hence, enable the network to provide a wider range of service quality levels and pricing policies.

## VI. EMOO OF RESOURCES IN PERSONALIZED WIRELESS NETWORKS

In this section, we present the building blocks of MOEAs utilized to solve the proposed OPA problem. For the purpose of this article, the implementation of the optimization formulation considers the optimization of the decision variable $\mathcal{X}_{u_b}$, whereas $\mathcal{P}_{u_b}$ is assigned a constant value for all users.

### A. SOLUTION ENCODING

As mentioned in Section V-B2, an OPA solution for one user is encoded as a binary vector $\mathcal{X}_{u_b}$, which represents a set of RBs available in the networks. With this in mind, the combined solution for all users is an $I_b$ x $N$ matrix in the form of

$$\mathcal{X}_{sol} = \begin{Bmatrix} x_{u_{(1,b)}}^{(1)} & x_{u_{(1,b)}}^{(2)} & \cdots & x_{u_{(1,b)}}^{(N)} \\ x_{u_{(2,b)}}^{(1)} & x_{u_{(2,b)}}^{(2)} & \cdots & x_{u_{(2,b)}}^{(N)} \\ \cdots & \cdots & \cdots & \cdots \\ x_{u_{(I_b,b)}}^{(1)} & x_{u_{(I_b,b)}}^{(2)} & \cdots & x_{u_{(I_b,b)}}^{(N)} \end{Bmatrix}. \tag{11}$$

A solution $\mathcal{X}_{u_b}$ is feasible if it meets the constraints associated with both objectives (9a) and (10a).

### B. OBJECTIVE FUNCTIONS

The proposed OPA problem in Section V-B4 is a bi-objective optimization problem with two objectives (1) average $\Delta_{\mathcal{U}_b}$ ($\overline{\Delta_{\mathcal{U}_b}}$) in (9a) and (2) average satisfaction in (10a). Each solution is evaluated in terms of the aforementioned contradicting objectives. In other words, maximizing user satisfaction will require lower $\Delta$s; hence, minimum resource-saving, and vice versa. The final Pareto front solutions trade-off these contradicting objectives.

### C. POPULATION INTIALIZATION

The population consists of $M$ solutions in the form of $I_b$ x $N$ matrixes. The initial population is generated by drawing the elements of the solution matrixes $\mathcal{X}_{sol}$ from a random binary uniform distribution.

### D. SELECTED MOEAs

In this article, we investigate the performance of five MOEAs in solving OPA. One of the most famous genetic algorithms considered in this article is non-dominated sorting evolutionary algorithm II (NSGA-II) [35] and its successor NSGA-III [36]. Besides, we investigate an indicator-based MOEA called $\varepsilon$-MOEA [37], [38]. Further, we consider SPEA2, which is a multiobjective evolutionary algorithm that incorporates the concept of elitism [38].

### E. EVOLUTIONARY OPERATORS

In this article, we utilize binary tournament selection as the selection operator for all algorithms [39]. As for crossover and mutation, we utilize the Half Uniform Crossover (HUX) operator and bit flip, respectively [40]. Also, the population size used across all experiments is 100 solutions.

### F. STOPING CRITERIA

In practice, wireless networks make decisions and perform actions in real-time; therefore, decision time is considered a crucial factor in solving OPA. In EMOO, decision time is proportional to the NFEs. Since OPA is a large scale optimization problem, practical systems are required to implement the appropriate techniques in order to meet the associated time constraint. Considering that meeting the wireless networks time constraint is out of the scope of this article, the stopping criterion for the implemented simulation is set to a predefined NFEs.

## VII. EMPIRICAL ANALYSIS AND DISCUSSION

In this section, we evaluate the performance of MOEAs described in the previous section in solving the formulated OPA problem. Then, the best performing MOEA is used to simulate personalized and non-personalized wireless networks in order to compare them in terms of the amount of saved resources and user satisfaction in the network. Also, we study the impact of errors and uncertainty introduced by the ML surrogate on the performance of MOEAs. Finally, we conduct several experiments to study the complexity and scalability of the proposed optimization problem.

The prototype and simulations in this article were done using Python 3.7.6. The DNN model was built using the TensorFlow library. In addition, the scikit-learn library was used for preprocessing the data, whereas seaborn and Matplotlib were used for visualization purposes. Also, MOO is performed using the Platypus library.

### A. EXPERIMENTAL SETTINGS
#### 1) CELLULAR NETWORK ENVIRONMENT

Consider a cell within a cellular network that covers Ottawa, Canada. The cell has one eNB and it is connected to users moving within its coverage area. The area of the cell is divided into a $k*k$ grid. The cellular network environment is simulated using the parameters listed in Table 3. The cellular network operator collects context data from users and stores it in a database. The collected data are of two types, real-time user satisfaction levels and context values. Measurements are recorded at each measuring instant. The period between two measuring instances is referred to as a Time Slot (TS). The service provider collects data from the users using a TS length of one second. Besides, the amount of resources consumed within each TS is recorded. Also, for the sake of simplicity, we assume that all users have the same minimum requirement for user satisfaction.

**TABLE 3.** Cellular network simulation parameters.

| Parameter name | Parameter value |
|---|---|
| Maximum number of available RBs | 100 |
| Number of subcarriers per RB | 12 |
| RB bandwidth ($B_{RB}$) | 180 kHz |
| Carrier frequency | 2 GHz |
| UE thermal noise density | -174 dBm/Hz |
| Grid size ($k$) | 100 |
| Flat fading | Rayleigh |
| Number of users ($I_b$) | 4 |
| Number of eNBs | 1 |
| $S_{\min, u_b}$ | 4 |
| $P^{\max}$ | 1 Watt |

#### 2) PERFORMANCE METRICS

The design of MOO metrics ususally considers three main performance criteria: capacity, convergence, and diversity [41]. Capacity metrics quantify the ratio (or number) of nondominated solutions in the solution space $\mathbb{S}$ that conforms to the predefined reference set. To measure MOO performance in terms of capacity, we calculate the Overall Non-dominated Vector Generation Ratio (NGR) [42]. NGR describes the capacity ratio of $\mathbb{S}$ with respect to $\mathbb{R}$, and is given as

$$NGR(\mathbb{S}, \mathbb{R}) = \frac{|\mathbb{S}|}{|\mathbb{R}|}, \quad (12)$$

where $|.|$ is the cardinality or number of elements in the set. In contrast to capacity, convergence metrics measures the proximity of the solution set $\mathbb{S}$ to the reference set $\mathbb{R}$. To measure MOO performance in terms of convergence, we calculate the Generational Distance (GD) [35] as follows:

$$GD(\mathbb{S}, \mathbb{R}) = \frac{(\sum_{i=1}^{|\mathbb{S}|} d_i^2)^{\frac{1}{2}}}{|\mathbb{S}|}, \quad (13)$$

where $d_i$ is the smallest distance from $s \in \mathbb{S}$ to the closest solution in $\mathbb{R}$, and is given as $d_i = \min_{r \in \mathbb{R}} ||F(s_i) - F(r)||$, where $s_i \in \mathbb{S}$. $||.||$ denotes the Euclidean distance and $F = (f_1(s), f_2(s))$, where $f_1$ is defined in (9a) and $f_2$ is defined in (10a). As for measuring the performance of MOO in terms of diversity, the Spacing (SP) metric [43] is calculated as follows:

$$SP(\mathbb{S}) = \sqrt{\sum_{i=1}^{|\mathbb{S}|} \frac{(d_i - \hbar)^2}{|\mathbb{S} - 1|}}. \quad (14)$$

In addition to the aforementioned metrics, we calculate the Hypervolume (HV) [44] and the Inverted Generational Distance (IGD) [45], [46]. HV and IGD measure the performance of MOO in terms of both convergence and diversity. HV is one of the most popular performance metrics for MOO, where it quantifies the volume in the objective space that is dominated by the solution set $\mathbb{S}$. HV is calculated as follows:

$$HV(\mathbb{S}, R) = volume(\bigcup_{i=1}^{|\mathbb{S}|} v_i), \quad (15)$$

where $v_i$ is the hypercube associated with $s_i \in \mathbb{S}$, and $R$ is a reference point. On the other hand, IGD is calculated as follows:

$$IGD(\mathbb{S}, |\mathbb{R}|) = \frac{(\sum_{i=1}^{|P|} d_i^2)^{\frac{1}{2}}}{|\mathbb{R}|}. \quad (16)$$

It is worth noting that the goal is to maximize HV and SP, whereas GD and IGD are better when they are minimized.

#### 3) REFERENCE SET GENERATION

In order to evaluate the performance and the quality of the Pareto front solutions, we need to compare them to a Reference set $\mathbb{R}$, which is the Pareto optimal set [34]. Since the optimal set is not known, we employ the common practice of using the best-known approximation of the Pareto optimal set as the reference set. The approximation of the optimal Pareto set for each instance is performed using the MOEAs listed in Section VI-D. Each MOEA is run 30 times and the final Pareto front solutions are merged into one set. Then, the set of merged Pareto front solutions is utilized to extract the set of non-dominated solutions, which represents the approximated Pareto front reference set $\mathbb{R}$. Using the simulation parameters in Table 3, we plot both objectives, the average $\Delta_{\mathcal{U}_b}$ ($\overline{\Delta_{\mathcal{U}_b}}$) vs. the average satisfaction for a single exemplary instance. For illustrative purposes, we set the minimum satisfaction allowed in the network to $S_{\min,u_b} = 1$. Fig. 6 illustrates the generated solution sets by NSGAII, NSGAIII, SPEA2, and $\varepsilon$-MOEA and their computed ranks. The Pareto front solutions are the set of solutions in the merged solution set that have the minimum rank (i.e., rank equal to zero), which are referred to as the non-dominated solutions set. In Fig. 7, the extracted optimum Pareto front solution set is plotted. The Pareto front solutions provide a trade-off between both objectives. Lower $\overline{\Delta_{\mathcal{U}_b}}$ values (i.e., a higher amount of consumed resources) offer higher user satisfaction levels in the network. To achieve a certain satisfaction level in the network, the optimum Pareto front solution for each instance is used to find the required minimum amount of resources. Depending on the service provider policy and the required $S_{\min,u_b}$, the personalized network will operate at the Pareto front solution point that achieves the required average satisfaction using the minimum resources. For the instance depicted in Fig. 7, if the required average satisfaction level for all users is 4, $\overline{\Delta_{\mathcal{U}_b}}$ should be less than or equal to 2.9, which is the maximum (i.e., optimum) $\overline{\Delta_{\mathcal{U}_b}}$ solution.

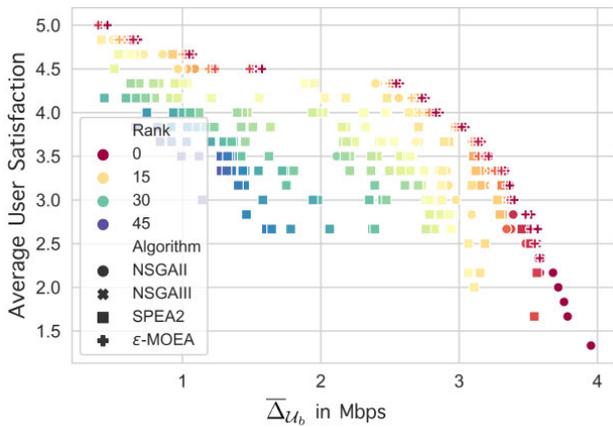

**FIGURE 6.** The generated approximation sets of the Pareto front solutions by the considered MOEA algorithms.

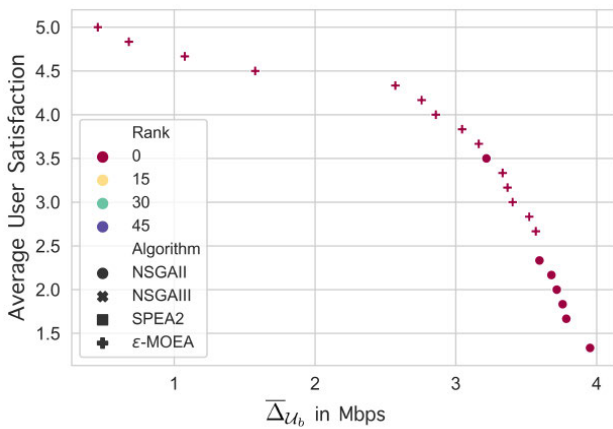

**FIGURE 7.** The non-dominated Pareto front reference solution set extracted from the multiple sets generated by the considered MOEA algorithms.

### B. EXPERIMENT 1: STATISTICAL ANALYSIS OF MOEAs PERFORMANCE RESULTS

In this experiment, we evaluate and compare the performance of the considered MOEAs in terms of the metrics described in Section VII-A2. Since some of the performance metrics use the reference set $\mathbb{R}$ as a parameter, an approximation of the reference set is computed for each instance. In order to test the significance of the performance results obtained in this section, we use non-parametric tests [47]. First, we use the Friedman $N \times N$ procedure to validate the existence of statistical differences among the results obtained by all algorithms. The Friedman test examines the null hypothesis ($H_0$) that the performance results for all algorithms come from the same distribution. In this article, we choose a significance level ($\alpha$) of 0.05. This means that if $p$–value is less than 0.05, the $H_0$ is rejected; hence, there exist statistical differences between the algorithms' performance results. Otherwise, $H_0$ cannot be rejected and the samples are likely coming from the same statistical distribution. If the Friedman test suggests the rejection of $H_0$, we perform several posthoc tests to examine the statistical difference of each algorithm from every other algorithm. The performed posthoc tests are Conover, Wilcoxon, Nemenyi, and Mann-Whitney [48].

In this experiment, to ensure the consistent performance of the chosen algorithm across all instances, we randomly choose a number of instances $N_m$ from the WPP dataset described in Section II-C. For each instance, we run MOEAs described in Section VI-D to solve the formulated OPA problem. Then, the performance metrics described in Section VII-A2 are computed. To ensure the statistical significance of the performance results for each instance, this process is repeated $N_s$ times for each instance in the selected $N_m$ instance set. Thereafter, the mean of each performance metric and for each algorithm is computed over all $N_s$ runs. The computed mean data is used to compare the algorithms.

#### 1) SAMPLE SIZE SUFFICIENCY
Before proceeding with the performance results comparison, it is important to determine whether the sample size is large enough to support our experiment. This has to do with the fact that too small sample size may produce inconclusive results. In literature, different sample sizes are used, yet a clear justification of the selection is rarely provided. One approach to choose a sufficient sample size $N_m$ and $N_s$ is to examine their relationship with the Standard Error of the Mean ($SE_M$), given as

$$SE_M = \frac{\sigma}{\sqrt{n}}, \qquad (17)$$

where $\sigma$ is the sample standard deviation and $n$ is the number of samples [49]. For the purpose of this article, we choose a maximum $SE_M$ of 0.05. In Fig. 8.a and Fig. 8.b, we compute and plot the $SE_M$ for the sample sizes $N_m$ and $N_s$, respectively. Using the $SE_M$ data in Fig. 8.a and Fig. 8.b, we choose the value of 30 samples for $N_s$ and $N_m$, which achieves $SE_M$ lower than 0.05.

**TABLE 4.** Statistical and Friedman test results.

| Metric | | NSGAII | NSGAIII | SPEA2 | ε-MOEA | p-value | Statistics |
|---|---|---|---|---|---|---|---|
| | | **Algorithms** | | | | | |
| HV | Rank | 3.520 | 3.880 | 1.970 | 4.580 | | |
| | Mean | 0.504 | 0.507 | 0.367 | 0.535 | < 0.001 | 336.184 |
| | Max | 0.721 | 0.723 | 0.629 | 0.722 | | |
| | Min | 0.213 | 0.199 | 0.012 | 0.266 | | |
| GD | Rank | 1.460 | 2.600 | 4.940 | 4.000 | | |
| | Mean | 0.013 | 0.019 | 0.098 | 0.038 | < 0.001 | 331.808 |
| | Max | 0.051 | 0.054 | 0.048 | 1.086 | | |
| | Min | 0.004 | 0.005 | 0.010 | 0.007 | | |
| IGD | Rank | 2.640 | 2.570 | 4.120 | 1.720 | | |
| | Mean | 0.221 | 0.217 | 0.318 | 0.192 | < 0.001 | 164.392 |
| | Max | 0.669 | 0.578 | 1.307 | 0.558 | | |
| | Min | 0.070 | 0.054 | 0.077 | 0.049 | | |
| SP | Rank | 3.120 | 2.080 | 3.985 | 4.810 | | |
| | Mean | 11.039 | 6.036 | 16.440 | 26.232 | < 0.001 | 363.668 |
| | Max | 19.728 | 16.882 | 33.966 | 69.815 | | |
| | Min | 4.739 | 1.184 | 0.000 | 3.267 | | |
| NGR | Rank | 3.980 | 2.145 | 5.000 | 2.875 | | |
| | Mean | 1.168 | 0.627 | 8.094 | 0.675 | < 0.001 | 388.476 |
| | Max | 2.539 | 0.873 | 15.940 | 0.910 | | |
| | Min | 0.716 | 2.590 | 0.077 | 9.206 | | |
| API | | -3.472 | 2.303 | -5.597 | **3.611** | | |

### 2) STATISTICAL ANALYSIS

Using the picked $N_s$ and $N_m$ values, we perform the described statistical analysis experiment. Table 4 summarizes the statistical and Friedman test results of the performed experiment. The Friedman test results show that the $p$–value for all performance metrics are less than $\alpha = 0.05$; hence, the test rejects $H_0$ and accepts the alternative hypothesis $H_a$. Consequently, Friedman test results suggest that, for each performance metric, there is a significant statistical difference among the metric values calculated for all algorithms.

In order to select the best performing algorithm, we employ the evaluated ranks by the Friedman test to compute a new metric, which we refer to as the Algorithm Performance Indicator (API), defined as follows:

$$\begin{aligned}API = &|w_{HV}|e^{j\theta_{HV}} * HV_r + |w_{SP}|e^{j\theta_{SP}} * SP_r \\&+ |w_{NGR}|e^{j\theta_{NGR}} * NGR_r + |w_{GD}|e^{j\theta_{GD}} * GD_r \\&+ |w_{IGD}|e^{j\theta_{IGD}} * IGD_r,\end{aligned} \qquad (18)$$

where $HV_r$, $NGR_r$, $GD_r$, $IGD_r$, $SP_r$ are the algorithm ranks of HV, NGR, GD, IGD, and SP. Besides, $|w_i|$ and $\theta_i$ $\forall i \in [HV, SP, NGR, GD, IGD]$ are the magnitude and phase of the weight $w_i$. The weights magnitude can be chosen based on the importance of each metric to the requirements of the tackled problem. Nonetheless, for the purpose of this article, we assign the same weight value of 1/5 to all weights; hence, all indicators are assumed to be of equal importance in the final score. As mentioned earlier, some metrics are better when their values are higher, and others are the opposite. To reflect this in our API score, we assign $\theta = 0$ for the metrics that are better when they are higher (i.e., HV, SP) and $\theta = \pi$ for those who are better when they are lower (i.e., GD, IGD). In comparison to other metrics, NGR could have $\theta = 0$ or $\pi$ depending on the value of $NGR_r$. This is because $NGR_r < 1$ is indicative of missing non-detected Pareto front solutions, whereas $NGR_r > 1$ is indicative of a higher number of detected solutions compared to the reference set solutions $\mathbb{R}$. Therefore, the closer NGR to 1, the better the quality of the Pareto front solutions. To incorporate this in our API equation, $\theta_{NGR}$ is defined by

$$\theta_{NGR} = (1 - \beta)\pi, \quad (19)$$

where $\beta$ is given by

$$\beta = \begin{cases} 0, & NGR > 1 \\ 1, & NGR \leq 1. \end{cases} \quad (20)$$

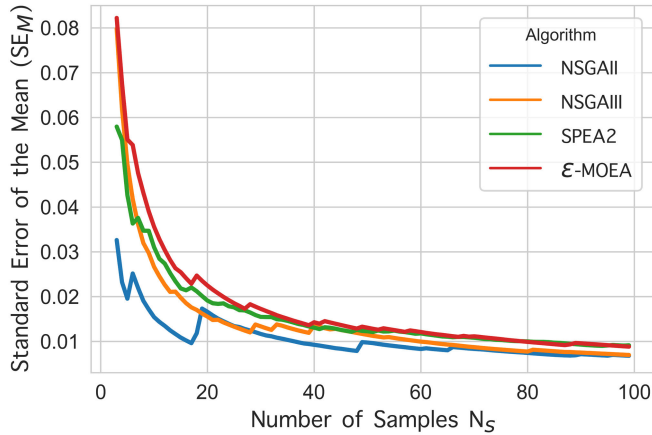

(a) $SE_M$ vs. sample size $N_s$.

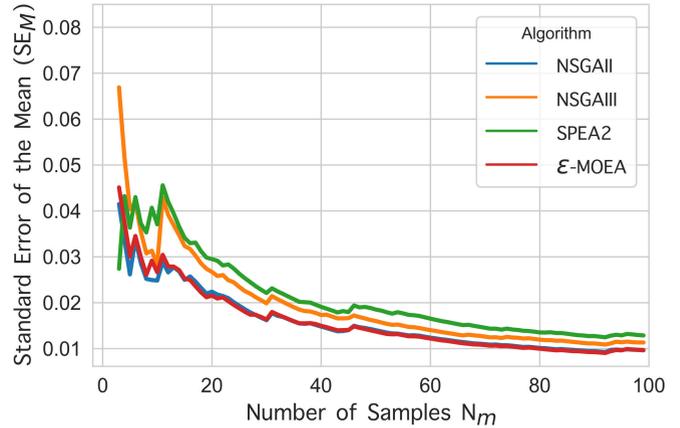

(b) $SE_M$ vs. sample size $N_m$.

**FIGURE 8.** Standard Error of the Mean ($SE_M$) vs. sample sizes.

Table 4 lists the API score for each algorithm. Of the four algorithms, $\varepsilon$-MOEA achieved the highest score of 3.611; therefore, we will use it for the personalized network simulations in the subsequent experiments.

Before accepting the aforementioned results, we should verify that the statistics for $\varepsilon$-MOEA are significantly different for the other algorithms. To test that, we performed the described pairwise posthoc tests. Table 5 summarizes the posthoc tests results. Although the posthoc tests failed to reject $H_0$ for some of the metrics computed for NSGAII and NSGAIII, $\varepsilon$-MOEA showed a significant statistical difference from every other algorithm.

### C. EXPERIMENT 2: PERSONALIZED VS. NON-PERSONALIZED WIRELESS NETWORKS

The goal of this experiment is to provide insights into the dynamics of personalized wireless networks and to show how they can be used to save the scarce network resources and improve user satisfaction levels in a controlled manner. Besides, the behavior of Surrogate-assisted Personalized Wireless Networks (SPN) is compared to Direct Feedback Personalized Wireless Networks (FPN) (i.e., networks utilizing direct user satisfaction feedback). Although the latter approach is not practical, we use it as a benchmark to study how the user satisfaction surrogates can deteriorate the optimum solutions of OPA, and consequently the amounts of savings and user satisfaction levels in the network. Also, both SPN and FPN are compared to the Non-Personalized Network (NPN), which tries to maximize the utilization of the available resources and maximize the provided rate. The wireless networks simulated in this section have four active users and NFEs is set to 5000 evaluations. Besides, the simulation time frame is set to 50 minutes. The networks described in this section are simulated at a resolution of one second (i.e., TS = 1 second); hence, the optimization of OPA is run at every second within the simulation time frame. Nonetheless, for visualization purposes, we average the results over every 30 seconds. Notably, the simulated SPN does not employ the surrogate management framework illustrated in Fig. 4.

**TABLE 5.** Pairwise comparison of algorithms rejected by the posthoc tests.

| Algorithms | HV | GD | IGD | SP | NGR |
|---|---|---|---|---|---|
| NSGAII vs. NSGAIII | | ✓ | | ✓ | ✓ |
| NSGAII vs. SPEA2 | ✓ | ✓ | ✓ | ✓ | ✓ |
| NSGAII vs. $\varepsilon$-MOEA | ✓ | ✓ | ✓ | ✓ | ✓ |
| NSGAIII vs. SPEA2 | ✓ | ✓ | ✓ | ✓ | ✓ |
| NSGAIII vs. $\varepsilon$-MOEA | ✓ | ✓ | ✓ | ✓ | ✓ |
| SPEA2 vs. $\varepsilon$-MOEA A | ✓ | ✓ | ✓ | ✓ | ✓ |

The first promised advantage of personalized networks is saving resources compared to current wireless networks.

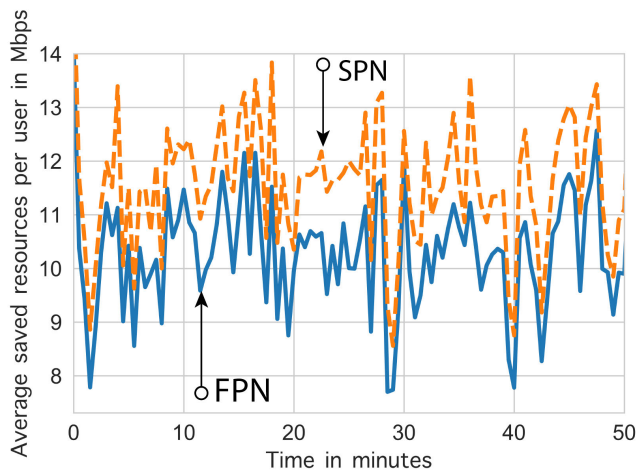

(a) The amount of saved resources ($QoS_{NP} - QoS_P$) by FPN and SPN for the four users vs. time in minutes.

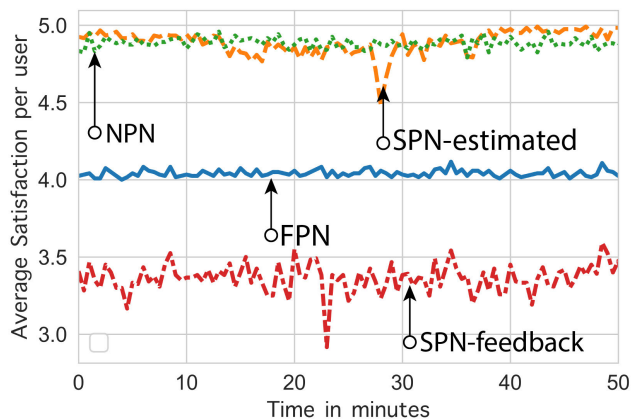

(b) Average user satisfaction for the four users vs. time in minutes for SPN and FPN.

**FIGURE 9.** The simulation results of the Direct Feedback Personalized Network (FPN) and the Surrogate-assisted Personalized Network (SPN) for four network users.

The amount of saved resources for both SPN and FPN is measured by $QoS_{NP} - QoS_P$, where $QoS_{NP}$ is the QoS provided by the non-personalized network and $QoS_P$ is the QoS provided by the personalized network, both in Megabits per seconds (Mbps). Fig. 9.a compares the amount of saved resources for FPN and SPN. The first thing we observe is the similar trends and patterns between both networks, which is indicative of the validity of the produced solutions by the surrogate-assisted OPA problem. Also, we further observe that the SPN spends fewer resources compared to the FPN; consequently, SPN achieved higher resource-savings. With this in mind, the amount of saved resources by the FPN is the maximum achievable amount of saved resources that doesn't compromise the required satisfaction levels required by the network operator. As mentioned earlier, another promised advantage of personalized networks is maintaining a specific user satisfaction level. In order to analyze the user satisfaction levels achieved by the simulated three networks, in Fig. 9.b we compare the average satisfaction levels for the four users vs. time for NPN, FPN. Also, for SPN, we plot the average satisfaction levels predicted by the surrogate model (SPN-estimated). In order to benchmark the predicted satisfaction results, we plot the actual satisfaction levels measured using direct user feedback (SPN-feedback). The first thing we observe from 9.b is that the satisfaction levels for NPN and FPN networks are above the specified minimum of $S_{\min, u_b} = 4$. Perhaps the most important observation from Fig. 9.b is the gap between the estimated (SPN-estimated) and the actual satisfaction levels for SPN (SPN-feedback). Although the SPN achieved superior amounts of saved resources compared to FPN (see Fig. 9.a), it failed to achieve the required average satisfaction level of 4. This is due to the satisfaction uncertainty introduced by the surrogate model, which led the SPN to further reduce resources below the minimum required to achieve $S_{\min, u_b} = 4$. These findings emphasize the importance of an effective surrogate management strategy to avoid the deterioration and divergence of user satisfaction levels resulting from false satisfaction predictions in the network.

### D. EXPERIMENT 3: THE IMPACT OF UNCERTAINTY INTRODUCED BY SURROGATES ON THE PERFORMANCE OF MOEAs

Generally, as shown in the previous experiment, the estimation error introduced by surrogates impacts the network's ability to use accurate user satisfaction information in the optimization process. The magnitude of this impact depends on several factors of which the most important is the performance of the utilized surrogate model. To further study this assumption, we design and perform the following experiment. In order to vary the performance level of the surrogate, we gradually increase the amount of training data. The accuracy and the amount of data used for training are recorded for each surrogate model. Then, using the set of trained surrogate models, we run the MOEAs to solve the OPA problem. For each surrogate model, each MOEA is run for 30 times; thereafter the average HV is computed. Fig. 10 compares the average HV values for the surrogate models with varying performance levels. Essentially, as shown in Fig. 10, as the quality of the employed surrogate model improves, the quality of the OPA solutions for all algorithms improve with different levels.

### E. EXPERIMENT 4: SCALABILITY ANALYSIS

In order to evaluate the scalability of the proposed formulation, we explore the effect of variables that contribute to the complexity of the problem. For our formulated problem, the number of users $\mathcal{U}_b$ determines the size of the problem decision variables; therefore, it increases the complexity of the problem. Another factor that impacts the complexity of the problem is the required quality of solutions. Higher quality solutions usually require higher NFEs; thereby, higher amounts of computing resources. In this section, we explore the effect of $\mathcal{U}_b$ and NFEs on complexity.

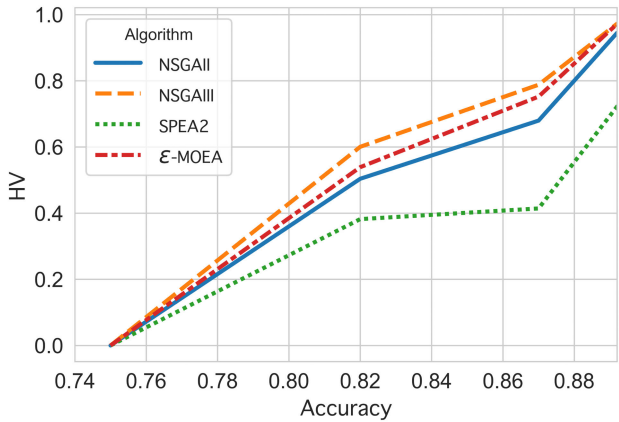

**FIGURE 10.** The average HV computed for different surrogate models with varying performance levels for NSGAII, NSGAIII, SPEA2, and $\varepsilon$-MOEA.

#### 1) THE IMPACT OF THE NUMBER OF NETWORK USERS ON COMPLEXITY

To study the effect of the number of network users $\mathcal{U}_b$ on OPA, we performed the following experiment. Using a random instance, the MOEAs considered in this article are run 30 times for varying number of users $\mathcal{U}_b$. NFEs is set to 5000 evaluations for each run. Then, HV is computed and averaged over the 30 runs for each $\mathcal{U}_b$. Fig. 11 depicts the averaged HV values vs. the number of users $\mathcal{U}_b$ for the considered MOEAs. From Fig. 11, we observe a descending HV trend as the number of users increases. Accordingly, we conclude the following: as the complexity of the problem increases, the quality of the output solutions decreases for a fixed amount of computing resources.

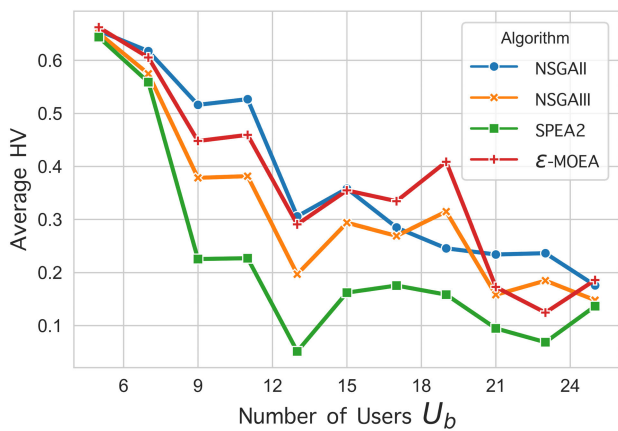

**FIGURE 11.** The number of users $\mathcal{U}_b$ vs. average HV for NSGAII, NSGAIII, SPEA2, and $\varepsilon$-MOEA.

#### 2) THE IMPACT OF THE REQUIRED SOLUTION QUALITY ON COMPLEXITY

In order to further improve the quality of the solutions as the complexity of the problem increases, MOEAs need a higher number of evaluations for each run; hence, more computing resources are required. In order to investigate this assumption, using a random instance, we compute the average HV with varying NFEs. In this experiment, the number of network users is fixed at 6 users. Fig. 12 compares the average HV vs. NFEs for the considered MOEAs. As anticipated, the HV values rise as the NFEs increase. After a certain NFEs limit, the average HV stagnates. In practice, the network should be able to decide the optimum NFEs in order to optimize the utilization of computing resources and make the computations more efficient. Usually, the optimum NFEs depends on several factors, including the network environment, the number of users, and the selected MOEA algorithm.

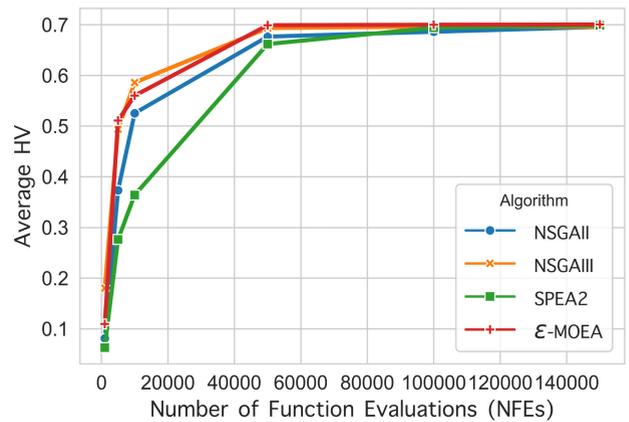

**FIGURE 12.** The Number of Function Evaluations (NFEs) vs. average HV for NSGAII, NSGAIII, SPEA2, and $\varepsilon$-MOEA.

### VIII. CONCLUSION

The complexity and the requirements of the emerging 5G and beyond applications make AI and ML an indispensable design tool for wireless networks. The ultimate design goal of any wireless network is to optimize two correlated and contradicting objectives, saving resources as well as maintaining high levels of user satisfaction. To this end, enabled by a big data-driven AI layer, wireless network personalization is proposed to optimize these two objectives, and thereby make fine-grained optimized decisions in networks. Due to the complexity and novelty of the proposed technology, several challenges need to be overcome. This article presented discussions on several design-related issues, including the integration of personalization into current wireless networks and modeling user satisfaction in wireless networks. The article focused on the decision-making process, which is part of the wireless network personalization framework. Particularly, the article proposed a MOO formulation to model the personalized resource allocation problem in wireless networks. The proposed MOO problem was tackled using evolutionary optimization due to its practicality and speed. Also, statistical analysis was conducted to verify the significance of the obtained results in this study. Using a dataset that represents a personalized wireless network environment, a simulation proof-of-concept prototype was built to solve the formulated problem. The prototype was utilized to demonstrate the benefits of implementing personalized networks in contrast to non-personalized networks. Also, the effect of uncertainty introduced by the ML surrogate models was examined. Lastly, a scalability analysis was performed to investigate the effect of increasing the number of problem variables, such as the number of users, on the complexity and quality of solutions.

### IX. FUTURE DIRECTIONS
This article addressed several challenges and issues that need to be overcome in order to realize wireless network personalization. Nevertheless, there are a number of critical open directions that are yet to be explored. In this section, we summarize and list some of these directions and open problems.

#### A. AN EFFECTIVE AND EFFICIENT SURROGATE MANAGEMENT SCHEME
In this article, we have proposed a surrogate management scheme in order to prevent the convergence of erroneous Pareto front solutions. Nonetheless, the prototype implemented in this article did not employ a surrogate management scheme; hence, the simulated surrogate-assisted personalized network was not able to maintain the required satisfaction level. Therefore, a more thorough study and analysis of the surrogate management scheme should be performed. Interestingly, the simulation results showed similar trends between the surrogate assisted (SPN-estimated) and direct feedback (SPN-feedback) personalized networks. Yet, the SPN-estimated had a negative bias in the satisfaction results and a positive bias in the saved resources results. For this reason, further analysis of the relationship between SPN-estimated and SPN-feedback is necessary. Furthermore, the simulation results presented in this article raised a vital question about the efficacy of adding a fixed bias to the outcome of surrogate-assisted personalized wireless networks in order to resolve the divergence issues and whether this bias can be calculated and predicted in real-time.

#### B. ACCURATE PREDICTION OF USER SATISFACTION
Among the main pillars of personalized wireless networks' design is the accurate prediction of user satisfaction levels. Therefore, further investigation into methods and advanced ML structures that can improve the accuracy of predicting user satisfaction in real-time is required. Besides, while the synthetic dataset utilized in this article provided valuable insights into the dynamics of personalized wireless networks, real-world user context and satisfaction data is fundamental to verify the developed models and framework.

#### C. ADVANED IEC MOEAs
As mentioned earlier, IEC MOEAs proposed in the literature are designed to incorporate user feedback into the optimization process. Leveraging IEC optimization in personalized networks could reveal drastic improvements in optimization performance and complexity reduction compared to traditional MOEAs.

#### D. SECURITY AND PRIVACY CONCERNS
The improved resource-saving and satisfaction levels promised by personalized networks are marred by privacy and security challenges. Personalized wireless networks entail gathering considerable amounts of data about users, such as context and satisfaction levels. For this reason, more research should be conducted to develop a responsible process to enable this technology without compromising users' privacy and confidentiality.

#### E. OPTIMIZING OTHER ASPECTS OF WIRELESS NETWORKS
This article highlighted resource scheduling, which is one of several aspects of wireless networks that can benefit from the proposed personalization scheme. Examples of such include network failure detection and network security decisions. Additionally, although the proposed OPA problem is formulated for cellular networks, it can be formulated and redesigned for any network with users, such as Wi-Fi and wired networks. Finally, the main ideas and the premise of the paper can be applied to any application that requires user feedback.

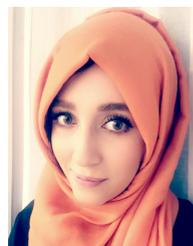

**RAWAN ALKURD** (Member, IEEE) is currently pursuing the Ph.D. degree with the Department of Systems and Computer Engineering, Carleton University, Ottawa, Canada. Her research interests include big data, artificial intelligence, and machine learning and their applications in wireless networks. In 2016, she received the Vanier Graduate Scholarship, Canada's most prestigious graduate scholarship.

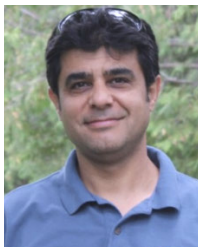

**IBRAHIM Y. ABUALHAOL** (Senior Member, IEEE) received the B.Sc., M.Sc., and Ph.D. degrees in electrical and computer engineering and the M.Eng. degree in technology innovation management. He is currently a Principal Data Scientist with Huawei Technologies and an Adjunct Research Professor with Carleton University, Ottawa, Canada. He is a Professional Engineer (P.Eng.) in Ontario, Canada. His research interests include machine learning and real-time big-data analytics and its applications in the Internet-of-Things, cybersecurity, and wireless communications.

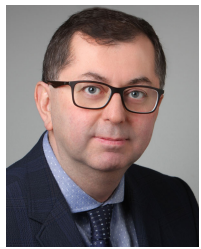

**HALIM YANIKOMEROGLU** (Fellow, IEEE) is currently a Full Professor with the Department of Systems and Computer Engineering, Carleton University. His collaborative research with industry has resulted in 37 granted patents. He is an Engineering with the Institute of Canada (EIC) and the Canadian Academy of Engineering (CAE). His research interest includes wireless technologies with special emphasis on wireless networks. He is a Distinguished Speaker for the IEEE Communications Society and the IEEE Vehicular Technology Society.

● ● ●